\documentclass[twocolumn,showpacs,preprintnumbers,amsmath,amssymb]{revtex4}
\usepackage{epsfig}
\newcommand{\bm}[1]{ \mbox{\boldmath $#1$}  } 

\begin{document}

\title{Energy distributions from three-body decaying many-body resonances}

\author{R. \'Alvarez-Rodr\'{\i}guez, A.S. Jensen, D.V. Fedorov, 
H.O.U. Fynbo }
\address{ Department of Physics and Astronomy, University of Aarhus, 
DK-8000 Aarhus C, Denmark}
\author{E. Garrido }
\address{Instituto de Estructura de la Materia, CSIC, Serrano 123, 
E-28006 Madrid, Spain}

\date{\today}

\begin{abstract}
We compute energy distributions of three particles emerging from
decaying many-body resonances.  We reproduce the measured energy
distributions from decays of two archetypal states chosen as the
lowest $0^{+}$ and $1^{+}$-resonances in $^{12}$C populated in
$\beta$-decays.  These states are dominated by sequential, through the
$^{8}$Be ground state, and direct decays, respectively.  These decay
mechanisms are reflected in the ``dynamic'' evolution from small,
cluster or shell-model states, to large distances, where the
coordinate or momentum space continuum wavefunctions are accurately
computed.
\end{abstract}


\maketitle

\paragraph*{Introduction.}

Energy and momentum conservation guarantees that two particles,
emerging from decay of a given quantum state, appear with definite
kinetic energies inversely proportional to their masses.  In
three-body decays the available energy can be continuously
distributed among the particles.  Prominent classical examples are
$\alpha$-emission and beta-decay, respectively.  Surprisingly enough
the decay of a many-body quantum system into three particles has not
been well described microscopically although discussed
phenomenologically for various systems.  The process depends on the
initial state and the dynamic evolution or equivalently the decay
mechanism.

This problem of three-body decay is common to several subfields of
physics. The invention of Dalitz plots was an early attempt to
classify the decay mechanisms by use of intermediate two-body doorway
states \cite{dal58}.  The underlying dynamics in particle physics may
be described as quark rearrangements. Similar decays occur in
annihilation of a proton-antiproton pair from a Coulomb-like orbit
into three mesons \cite{ppb88}.  In molecular physics an example is
decay of excited states of the H$_3$-molecule into three hydrogen
atoms \cite{gal04}.  In nuclear physics there exists a large number of
three-body decaying systems of disparate structures and decay
mechanisms, e.g. various excited states of $^6$He, $^6$Li, $^{12}$C,
$^{17}$Ne. More and more high-quality experimental data become
available in all subfields \cite{gal05,bla03,fyn03,dig05} and
quantitatively accurate models are needed to extract and understand
the underlying physics.

The purpose of the present letter is to compute the energy
distributions for three-body decaying excited nuclear many-body
resonances.  We shall assume that the resonances are populated in
$\beta$-decays and consequently only an outgoing flux is present.  For
reactions an ingoing flux is required to provide the population of the
decaying wavefunction.  Such a generalization is easily achieved by
allowing initial conditions different from that of a resonance
wavefunction.  In all cases the major difficulty is to compute
accurately the asymptotic large-distance three-body wavefunctions
corresponding to genuine many-body resonances, which possibly differ
at small distances from cluster states formed by the emerging three
particles \cite{gar07}.

At least four problems must be solved to overcome the difficulties,
i.e.  (i) the complex scaled resonance wavefunctions must be
accurately determined even though they vanish exponentially at large
distances, (ii) the wavefunctions must be traced as they ``evolve
dynamically'' from relatively small to asymptotically large distances,
(iii) the Coulomb problem of coupling continuum states at infinitely
large distances must be solved, (iv) the mixture of two- and
three-body asymptotics must be accurately determined.

\paragraph*{Theoretical framework.}

We use the hyperspherical adiabatic expansion method of the Faddeev
equations combined with complex scaling \cite{nie01,fed02}.  The
hyperradius $\rho$ is the most important of the coordinates.  For
three identical particles of mass $m_{\alpha}$ the definition is
\begin{eqnarray}  
  m_N  \rho^{2} =  \frac{m_{\alpha}}{3} \sum^{3}_{i<j}
 \left(\bm{r}_{i}-\bm{r}_j\right)^{2} = 
  m_{\alpha}  \sum^{3}_{i=1} 
 \left(\bm{r}_{i}-\bm{R}\right)^{2}  \label{e120} \;,
\end{eqnarray}
where $\bm{r}_{i}$ is the coordinate of particle number $i$, $\bm{R}$
is the three-body center-of-mass coordinate and $m_N = m_{\alpha}/4$.

The two-body interactions are chosen to reproduce the available
low-energy scattering data.  A three-body potential with a range
corresponding roughly to the three touching constituent particles is
adjusted to reproduce the energy position of the many-body resonance
under investigation.  The wavefunction and the complex energy of the
resonance are then defined. Coordinate and momentum space angular
wavefunctions are identical for a given total energy and an
asymptotically large value of $\rho$ \cite{gar05b}.  Finally, the
energy ($E_{\alpha}$) distribution of the particle is obtained by
integrating the square of the resonance wavefunction over unobserved
momenta.

The numerical computations must first provide accurate wavefunctions
from small to intermediate distances, where the relatively fast
changes due to the crucial short-range interaction are completed. This
is efficiently achieved with the Faddeev decomposition, and a
hyperspherical harmonics basis size individually adjusted to the
accuracy needed for the different partial waves.  The smoother
variation from intermediate to asymptotic distances is analytical for
short-range interactions \cite{nie01}.  The long-range Coulomb
potentials can be treated numerically precisely as any other potential
in the transitions from small to intermediate distances.

\begin{figure}[h]
\begin{center}
\vspace*{-1.5cm}
\epsfig{file=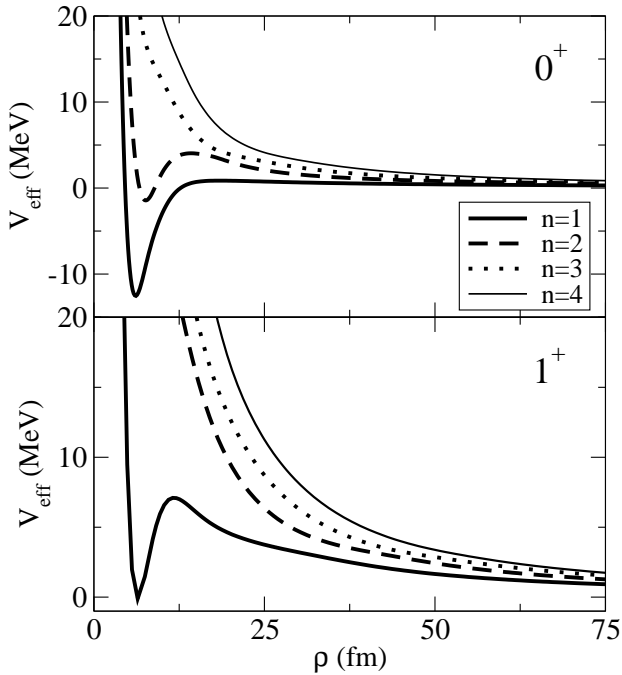,scale=0.45,angle=0}
\end{center}
\vspace*{-2.6cm}
\caption{The real parts of the four lowest adiabatic effective 
potentials, including the three-body potentials, as functions of
$\rho$ for the $0^{+}$ and $1^{+}$ resonances of $^{12}$C.  The
two-body interaction, obtained from \cite{fed96}, is a slightly
modified version of the a1-potential of \cite{ali66}.  The parameters
of the three-body Gaussian potentials, $S \exp(-\rho^2/b^2)$, are
$b=6$~fm and $-S =20,92$~MeV for $0^{+}$ and $1^{+}$, respectively
\cite{alv07}. }
\label{fig1}
\end{figure}

However, special treatment is required in the variations from
intermediate to large distances.  The necessary basis sizes become
insurmountably large.  Our solution is to compute an accurate
wavefunction at intermediate, but relatively large, distances.  This
is achieved when an even larger hyperradius, compensated by a larger
basis, leads to the same observables derived from the wavefunctions.
This stability condition is difficult to reach when both three-body
background continuum states are populated simultaneously with
resonances in one or more of the two-body subsystems.  At sufficiently
large distances we can precisely identify these structures as
components in the complex scaled wavefunctions related to different
adiabatic potentials \cite{gar05b}, e.g. sequential decay proceeds
through a potential approaching the corresponding complex two-body
energy $E_{2b}$ \cite{fed02}, whereas no intermediate structure is
present for direct decay to the continuum.

When the two-body intermediate states have large widths the related
radial wavefunction decreases quickly, because then the adiabatic
couplings are large. These states then dissipate fast into the
continuum described as direct decay and the distinction becomes
artificial. This process eventually happens for all sequential decays
since the intermediate states are unstable.  Classification into
sequential and direct decay is related to the use of different
complete basis states, i.e. either two-body resonances and the third
particle in the continuum or three-body continuum states.
Interpretation as sequential or direct is then meaningful when a few
states in one basis are sufficient while many are needed in the other.
Such a distinction between paths producing the same observable is not
possible in quantum mechanics.

\paragraph*{Illuminating archetypes.}

Accurate measurements of $\alpha$-particle energy distributions are
available from decays of $0^{+}$ and $1^{+}$-resonances of $^{12}$C
\cite{fre94,fyn03,dig05,dig06}. The lowest $0^{+}$-resonance is often
described as a cluster state \cite{kan98,nef04,des02,pie05,nav00}
whereas the $1^{+}$-resonance in contrast is referred to as a
shell-model state without any significant cluster structure
\cite{kan98,nef04,des02,pie05,nav00}.  Furthermore, the decay
mechanisms are known to be different~\cite{fyn03,dig05}.  These cases
are therefore ideally suited as illustrations of the present novel
technique.

\begin{figure}[h]
\vspace*{-0.5cm}
\begin{center}
\epsfig{file=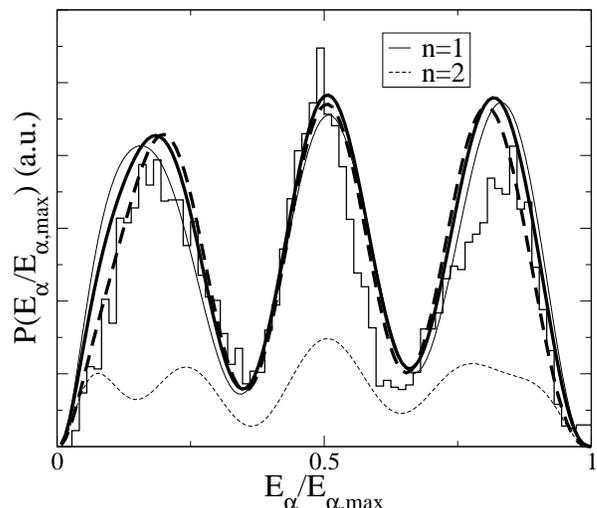,scale=0.34,angle=270}
\end{center}
\vspace*{-1.0cm}
\caption{ The $\alpha$-particle energy distribution for the $1^{+}$
resonance of $^{12}$C at 5.43~MeV above threshold at an excitation of
12.71~MeV.  The energy is measured in units of the maximum possible,
i.e. $2\times 5.43 /3$~MeV.  The interactions are given in
fig.~\ref{fig1}.  The thick solid and dashed curves are for coordinate
space wavefunctions at $\rho=70,100$~fm. The thin curves are
contributions from separate adiabatic potentials.  The momentum space
computation (described below) fall on top of the $\rho=100$~fm curve
(thick dashed).  The histogram is the experimental distribution
\cite{dig06}. }
\label{fig2}
\end{figure}

In fig.~\ref{fig1} we show the lowest potentials where the attractive
pockets at small distance support the resonances and provide the small
distance boundary conditions. As the hyperradius increases beyond the
barriers, the potentials all decrease as $1/\rho$ due to the Coulomb
repulsion.  The structure at large distances is necessarily of
three-body character since this is the boundary condition imposed by
the measurement. In contrast, at small distances these clusters
overlap and the detailed description must use the nucleon degrees of
freedom.  The first adiabatic potential corresponds to the
$^{8}$Be($0^+$) state and therefore associated with this sequential
decay.  We shall explore the, perhaps surprising, conjecture that the
decay can be described almost entirely within the present cluster
model.

\paragraph*{The $1^{+}$-resonance.}

With the potentials in fig.~\ref{fig1} we show the computed energy
distribution in fig.~\ref{fig2} for the $1^{+}$-resonance where
sequential decay via the $^{8}$Be ground-state is forbidden.  The
asymptotic behavior is reached already for hyperradii larger than
about 60~fm. The small variation of the distribution from 70~fm to
100~fm show the convergence and the stability.  Higher accuracy is
obtained at these distances with a moderate basis size than at larger
distances where the basis quickly becomes insufficient.  Two
interfering adiabatic potentials are necessary to reach the impressive
agreement with the measured distributions.  It is remarkable that the
cluster model provides this accuracy in spite of the fact that the
initial decaying state is a many-body resonance without any three-body
structure.

\begin{figure}[h]
\begin{center}
\vspace*{-1.7cm}
\epsfig{file=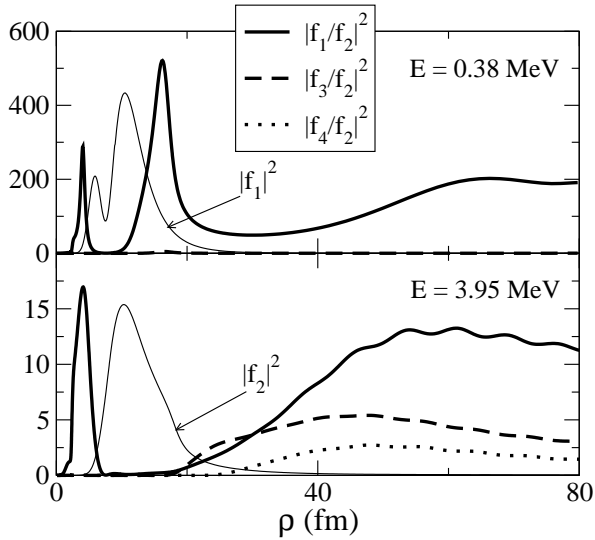,scale=0.46,angle=0}
\end{center}
\vspace*{-4.6cm}
\caption{The four lowest radial wavefunctions as functions of $\rho$ 
for each of the two $0^{+}$-resonances of $^{12}$C at $0.38$~MeV and
3.95~MeV above threshold or at excitation energies of $7.63$~MeV and
11.2~MeV \cite{dig06}.  Ratios and the small-distance dominating
wavefunctions are given for both states. }
\label{fig3}
\end{figure}

To test the reliability we Fourier transformed the wavefunction in two
ways, i.e. first numerically with coordinates from $\rho=0$ to
$100$~fm and secondly by use of the analytic solution obtained from
the parametrized adiabatic potentials which asymptotically are sums of
$1/\rho$ and $1/\rho^2$ terms.  The results are remarkably similar
distributions and the analytic result is in fact indistinguishable
from the curve for $\rho=100$~fm in fig.~\ref{fig2}.  Small deviations
from the measurements could be due to two-body interactions with
resonance properties deviating slightly from the values measured for
$^{8}$Be.  However, most of the differences are more likely due to
uncertainties arising from acceptance of the detectors used in the
experiment.

It is amusing to estimate that the sequential decay via the $^{8}$Be
$2^{+}$-resonance would produce a similar central peak of about twice
the width. To reproduce the data strong interference would then be
necessary.  Good phenomenological reproduction of the data is obtained
by R-matrix theory where the smaller width is explained due to
preferentially populating the low energy tail of the $^{8}$Be
$2^{+}$-resonance, and where effects of interference also play an
important role ~\cite{fyn03}.

\begin{figure}[h]
\begin{center}
\vspace*{-1.5cm}
\epsfig{file=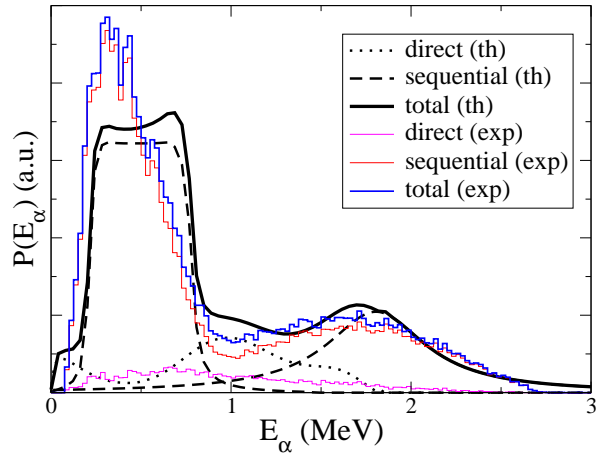,scale=0.34,angle=270}
\end{center}
\vspace*{-0.8cm}
\caption{The $\alpha$-particle energy distribution for the second
$0^{+}$ resonance of $^{12}$C at 4.3~MeV above threshold at an
excitation of 11.2~MeV \cite{dig05,dig06}.  The interactions in
fig.~\ref{fig1} give an energy of 3.95~MeV above threshold
\cite{alv07}.  We use $2.8$~MeV to account for interference and 
beta-feeding distortion.  The maximum energy for the most likely
resonance position is then $2\times 2.8 /3 \approx 1.8$~MeV.  The
histograms are experimental direct (the small part), sequential and
total distribution.   }
\label{fig4}
\end{figure}

\paragraph*{The $0^{+}$-resonances.}

The complex scaled radial wavefunctions are shown in fig.~\ref{fig3}
for the two lowest $0^{+}$-resonances. The largest probability is
found at small distance and they all vanish with increasing
hyperradius. Their relative sizes are fairly insensitive to variations
of the hyperradius at large distances where the energy distributions
are determined.  The first resonance is described by the first
adiabatic component for all distances whereas the second resonance
changes character from small to large $\rho$ from 4\% to 75\% of the
first adiabatic component.  This component, which dominates for both
resonances at large distance, approaches the configuration of the
$0^{+}$-resonance in $^{8}$Be with the third $\alpha$-particle far
away, i.e. $s$-waves between each pair of $\alpha$-particles for each
of the three Faddeev components.  The result is an energy distribution
with characteristic features of sequential three-$\alpha$ decay,
i.e. a narrow high-energy peak and a distribution around one quarter
of the maximum of the $^{12}$C resonance energy.  Unfortunately these
computed distributions are not accurate because the two-body
asymptotic behavior in these cases are not reached for $\rho$ less
than 100~fm.  However, the method provides the amount of sequential
decay and we can substitute the inaccurate component by the known
two-body asymptotic behavior. The energy distribution from decay of
the first $0^{+}$-resonance at $0.38$~MeV is then seen from
fig.~\ref{fig3} to be almost entirely determined by the first
potential which means sequential decay.  The direct decay is only
about 1\% in agreement with the experimental upper limit
\cite{fre94}. This energy distribution is then in complete agreement
with experimental data.

The second $0^{+}$-resonance is also dominated by the first adiabatic
potential at large distance.  This is in striking contrast to the
domination by the second potential at small distance.  This is an
example of the importance of the dynamical evolution from small to
large distances.  The result is about 75\% sequential (first
potential) and 25\% direct decay described by the other adiabatic
potentials.  In comparison with measurements complications arise for
two reasons both related to the large width of the order $1$~MeV.
First, effects of energy-dependent feeding in the beta-decay
populating the decaying state are substantial in the data
\cite{dig05,dig06}.  Higher beta-energies are rather strongly favored
resulting in distributions moving towards lower energies.  Second, the
experimental analysis is hampered by possible effects from other
resonances.  Their contributions are possibly not fully disentangled.

The peak energy corresponding to the resonance position is at about
$2.8$~MeV in the beta-feeding process \cite{dig05,dig06}.  We
illustrate in fig.~\ref{fig4} the sequential part of the energy
distribution of the first emitted $\alpha$-particle by using the
Breit-Wigner distribution defined with the most probable position at
$2.8$~MeV and a width equal to the sum of the widths of the three-body
decaying resonance and the intermediate two-body resonance.  The two
$\alpha$-particles following from decay of $^{8}$Be are uniquely
related by kinematic conditions resulting in a peak at lower energy.
The large width of the three-body decaying resonance smears out the
latter distribution.  Between these two peaks appears the contribution
of about 25\% from direct decay described by the other adiabatic
potentials.  The inaccuracies in the computed distributions are first
that deviations from the Breit-Wigner shape become important for the
large width of $1$~MeV, and second that the fraction of sequential
decay may be underestimated by perhaps 10\% due to missing higher
partial waves.

In any case, the shape of the sequential decay via the $^{8}$Be ground
state is derived by precisely the same kinematic conditions in both
the computation and the experimental analysis.  The largest
differences between theory and experiment is simulated by the shift of
resonance peak energy.  The agreement is rather good and only possible
due to the computed decay mechanism of dynamic evolution with
hyperradius.

\paragraph*{Summary and conclusions.}

We have computed the energy distributions for three-body decaying
many-body resonances.  Combinations of short-range and repulsive
Coulomb interactions are allowed. We conjecture, and show in specific
cases, that the energy distributions of the decay fragments are
insensitive to the short-distance many-body structure, but accessible
in a three-body cluster model.  The resonance structures may be
completely different at small and large distances. This dynamic
evolution is decisive for the decay mechanism.  We separate components
with two- and three-body asymptotics corresponding to sequential and
direct decays.  This distinction is crucial to obtain accurate
wavefunctions at large distances.  We test the method by comparing
results from coordinate and momentum space.

We illustrate by application to the archetypes of $\alpha$-decaying
$0^{+}$ and $1^{+}$-states in $^{12}$C. The $1^{+}$-resonance cannot
be described as a three-body state but its decay proceeds directly
into the three-body continuum.  The two $0^{+}$-resonances both have
substantial, but very different, cluster components at small
distances.  However, they both decay preferentially through the same
large distance structure best described as the $0^{+}$-resonance of
$^{8}$Be. These sequential decays imply a total rearrangement of the
second of these resonances from small to large distances.  The
accurately measured $\alpha$-particle energy distributions for all
three resonances populated in $\beta$-decay are reproduced remarkably
well.  Thus the method has past very severe tests.  It is reliable and
with predictive powers.

\paragraph*{Acknowledgments.} 

R.A.R. acknowledges support by a post-doctoral fellowship from
Ministerio de Educaci\'on y Ciencia (Spain).


\vspace*{-0.3cm}



\begin{thebibliography}{99}


\bibitem{dal58} R.H. Dalitz, Philosophical. mag. {\bf 44}, 1068 (1953).  

\bibitem{ppb88} C. Amsler, Rev. Mod. Phys. {\bf 70}, 1293 (1998). 

\bibitem{gal04} U. Galster, U.  M\"{u}ller, H. Helm, 
Phys. Rev. Lett. {\bf 92}, 073002 (2004).

\bibitem{gal05} U. Galster, F. Baumgartner, U.  M\"{u}ller, H. Helm, 
and M.Jungen,  Phys. Rev. {\bf A72}, 062506 (2005). 

\bibitem{bla03} B. Blank {\it et al.}, C. R. Physique {\bf 4}, 521 (2003).

\bibitem{fyn03} H.O.U. Fynbo et al., Phys. Rev. Lett. {\bf 91}, 082502 (2003).

\bibitem{dig05} C. Aa. Diget et al., Nucl. Phys. {\bf A 760}, 3 (2005).

\bibitem{gar07} E. Garrido, D.V. Fedorov, H.O.U. Fynbo and A.S. Jensen, 
Nucl. Phys. {\bf A 781}, 387 (2007).

\bibitem{nie01} E. Nielsen, D.V. Fedorov, A.S. Jensen, and E. Garrido,
Phys. Rep. {\bf 347}, 373 (2001).

\bibitem{gar05b} E. Garrido, D.V. Fedorov, A.S. Jensen and H.O.U. Fynbo,
Nucl. Phys. {\bf A 766}, 74 (2005).

\bibitem{fed02}  D.V. Fedorov, E. Garrido, and A.S. Jensen,
Few-body systems, {\bf 33}, 153 (2003).

\bibitem{fre94} M. Freer {\it et al.}, Phys. Rev. {\bf C49}, R1751 (1994) 

\bibitem{dig06}  C. Aa. Diget, PhD-thesis, IFA,
Univ. of Aarhus, 2006. 

\bibitem{kan98}  Y. Kanada-En'yo, Phys. Rev. Lett. {\bf 81}, 5291 (1998).

\bibitem{nef04} T. Neff and H. Feldmeier,
Nucl. Phys.{\bf A 738}, 357 (2004).

\bibitem{des02} P. Descouvemont, Nucl. Phys.{\bf A 709}, 275 (2002).

\bibitem{pie05}  S.C. Pieper, Nucl. Phys.{\bf A 751}, 516 (2005).

\bibitem{nav00} P. Navratil,  J.P. Vary, and B.R. Barrett,
Phys. Rev. {\bf C 62}, 054311 (2000).

\bibitem{fed96}  D.~V.~Fedorov and A.~S.~Jensen,
Phys.~Lett. {\bf B389}, 631  (1996).

\bibitem{ali66}  S. Ali and A.R. Bodmer,
Nucl. Phys. {\bf 80}, 99 (1966).

\bibitem{alv07} R. \'Alvarez-Rodr\'{\i}guez, E. Garrido,  A.S. Jensen,  
D.V. Fedorov, and H.O.U. Fynbo, Eur. Phys. J. {\bf A 31}, 303 (2007).  







\end{thebibliography}
\end{document}